# Day X:  Ultramafic talc-carbonate unit

## The North Pole Dome and Dresser Formation

Adrian Brown, Plancius Research

The North Pole Dome (NPD) is located in the centre of the East Pilbara Terrane (Van Kranendonk et al., 2007). The NPD is a structural dome of bedded, dominantly mafic volcanic rocks of the Warrawoona and Kelly Groups that dip gently away from the North Pole Monzogranite exposed in the core of the dome (Figure 1) (Van Kranendonk, 1999, 2000). Average dips vary from 30 to 60° in the inner part of the dome to about 60 to 80° in the outer part of the dome (Van Kranendonk, 2000). The North Pole Monzogranite is interpreted to represent a syn-volcanic laccolith to the Panorama Formation (Thorpe et al., 1992) and has been estimated to extend approximately 1.5km below the surface, based on gravity surveys (Blewett et al., 2004). Felsic volcanic formations are interbedded with the greenstones (Hickman, 1983), and these are capped by cherts that indicate hiatuses in volcanism (Barley, 1993; Van Kranendonk, 2006). An overall arc-related model for hydrothermal activity is favored by Barley (1993), whereas more recent studies have indicated a mantle-plume model for igneous and hydrothermal activity at the North Pole Dome (Van Kranendonk et al., 2002, 2007; Smithies et al., 2003; Van Kranendonk and Pirajno, 2004).

## North Pole Dome Hyperspectral Spectroscopy Project

Airborne Hyperspectral Reflectance Spectroscopy Imaging (here called 'hyperspectral imaging') is the practice of using remote reflectance spectroscopy to construct maps of terrestrial target sites. Hyperspectral imaging may be viewed as a natural extension of aerial photography. A camera records light in just one part of the visible (0.4–0.7μm) region of the electromagnetic spectrum (EM). Hyperspectral imaging records the intensity of radiation arriving at the sensor at many different wavelengths across the EM spectrum. Each measurement at a different wavelength constitutes a separate spectral channel. (A channel is sometimes called a band; here the term channel is preferred in order to prevent confusion with the term 'absorption band'.) Each channel may be viewed as a separate greyscale image, which when stacked together form a three dimensional image cube.

Hyperspectral imaging utilises the phenomenon of anomalous absorption of photons by molecules to discriminate between different minerals. Crystalline minerals (those with a repeating pattern of atoms) absorb light (photons) of a characteristic energy (and consequently wavelength), and therefore minerals of a known structure can be recognised by characteristic absorption bands in their reflectance spectrum (Hunt, 1979).

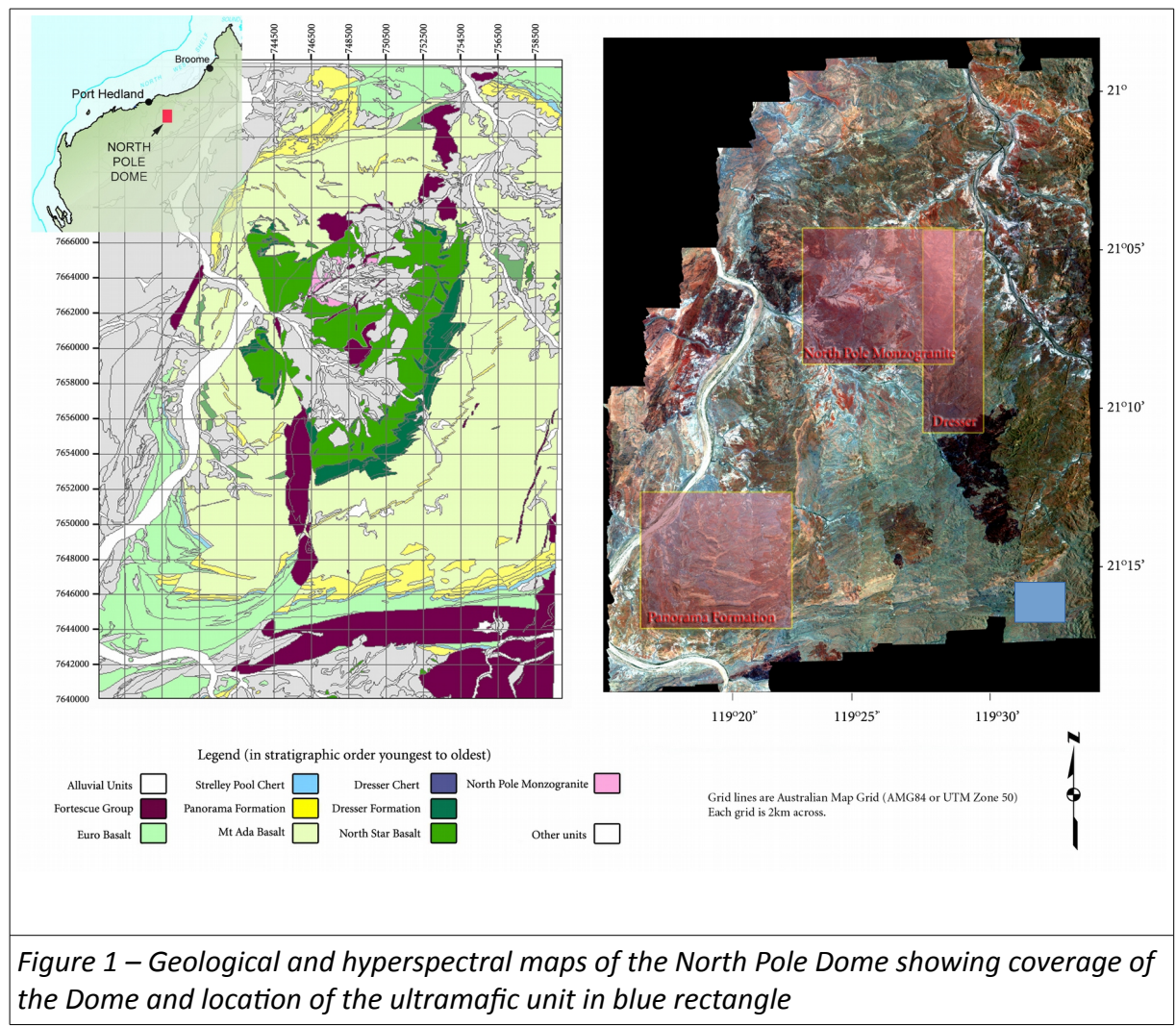

*Figure 1 – Geological and hyperspectral maps of the North Pole Dome showing coverage of the Dome and location of the ultramafic unit in blue rectangle*

The hyperspectral dataset analysed in the study consists of 14 swathes, 2.3 km wide on average, and from 6 to 22 km long (Figure 1). It was collected on October 22 2002, from 1030–1230 Australian Western Standard (local) Time (AWST). The ground instantaneous field of view (or pixel size) was approximately 5 m in the cross-track and down-track dimensions. The collection date was at the end of the tropical dry season in order to minimize vegetation coverage. The swathes were flown in order from east to west, odd swathes flying south, even swathes in a north direction. The average altitude of the aircraft was 2460 m. The mean height above sea level for each swathe varied from 46 to 85 m, with an overall average elevation of 65 m. The latitude and longitude of the NE corner of the dataset is approximately 21°S, 119.5°E (Figure 1).

The dataset was provided by the Australian company HyVista, the operator of the HyMap instrument, in 'radiance-at-sensor' form. It was corrected for atmospheric effects using the HyCorr computer program, provided by CSIRO, which is similar to the ATREM algorithm (Gao, 1993).

### Apex Basalt Olivine cumulate

The rocks of the Warrawoona Group constitute two komatiitic- thoelitic-felsic-chert volcanic successions which have ages spanning 3.515-3.426 Ga - as each succession gets younger in age, it gets progressively less mafic (Van Kranendonk et al., 2002). The hyperspectral VNIR signature of talc has been used to map a komatiite layer around the North Pole Dome in the Apex Basalt member of the Warrawoona Group (Brown et al., 2004). The Apex Basalt overlies the stromatolite-bearing 3.49 Ga Dresser Formation chert-barite unit, which probably represents the late stage of an active volcanic caldera (Van Kranendonk et al., 2008). The komatiitic 3.46 Ga Apex Basalt probably represents resumption of distal volcanic activity following a ~ 20 k year hiatus. Talc-carbonate hydrothermal alteration of the Apex Basalt was either achieved on emplacement of the komatiite or when the overlying theolitic 3.46 Ga Mt. Ada Basalt unit was emplaced.

### Komatiite lavas

Komatiite lavas form when high-temperature (~1400-1600 °C), low viscosity (0.1-1 Pa), mantle derived, ultramafic lavas are extruded and flow turbulently at the surface. Komatiites are found almost exclusively in Archean shield areas due to the higher heat of the Earth's mantle during that period (Campbell et al., 1989). Treiman (2005) has convincingly demonstrated that the Nakhalite meteorites have a similar texture to komatiite cumulate layers in the Canadian Archean. Komatiite lavas have also previously been proposed as possible analogs for Martian rocks on geochemical (Baird and Clark, 1981), morphological (Reyes and Christensen, 1994; Rampey and Harvey, 2012) and microtextural (Treiman, 2005) grounds. The komatiite layer detected in the North Pole Dome was associated with talc-carbonate alteration which has been hypothesized to be the result of hydrothermal alteration (Brown et al., 2005, 2010).

*Thin section and outcrop*

Figures 3a and 3b show the olviine cumulate layer in outcrop and in thin section, as reported in Brown (2006). The thin section shows that the olivine has been replaced by talc and serpentine in these two locations. These figures indicate that the olivine cumulate rocks tend to accumulate in mounds that weather characteristically.

| | | | |
|---|---|---|---|
| Sample ID | AJB0504100 | 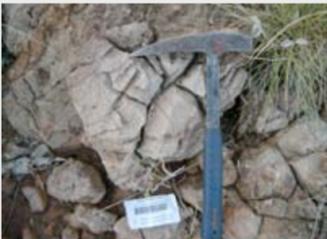 | 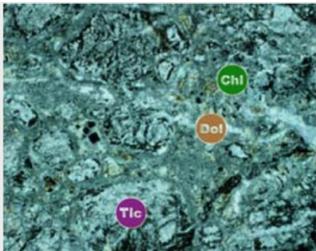 |
| Location | 754702E 7663703N | | |
| Unit | AWak | | |
| Description | Carbonated ultramafic | | |
| Alteration Zone | TALC-CARBONATE PROPYLITIC | | **PTS Summary:** Orbicular blebs of talc replacing olivine surrounded by carbonate in veins. XPL. |

Figure 3a. Olivine cumulate replaced by talc – note mounding of subcrop.

| | | | |
|---|---|---|---|
| Sample ID | AJB0503098 | 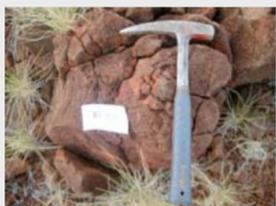 | 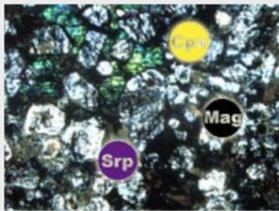 |
| Location | 754679E 7663523N | | |
| Unit | AWak | | |
| Description | Peridotitic komatiite | | |
| Alteration Zone | SERPENTINITIC PROPYLITIC | | **PTS Summary:** Cumulate serpentine after olivine with interstitial (green) pyroxene and magnetite (black). XPL. |

Figure 3b. Olivine cumulate showing replacement by serpentine – again, note distinctive mounding of the subcrop.

Figure 4 shows a HiRISE image of the olivine-carbonate lithology at Jezero crater, this shows the light toned floor of Jezero crater overlain by the delta deposit. Note the characteristic polygonal fracturing of the light toned floor unit, which is similar to the mounds of the Apex Basalt cumulate layer.

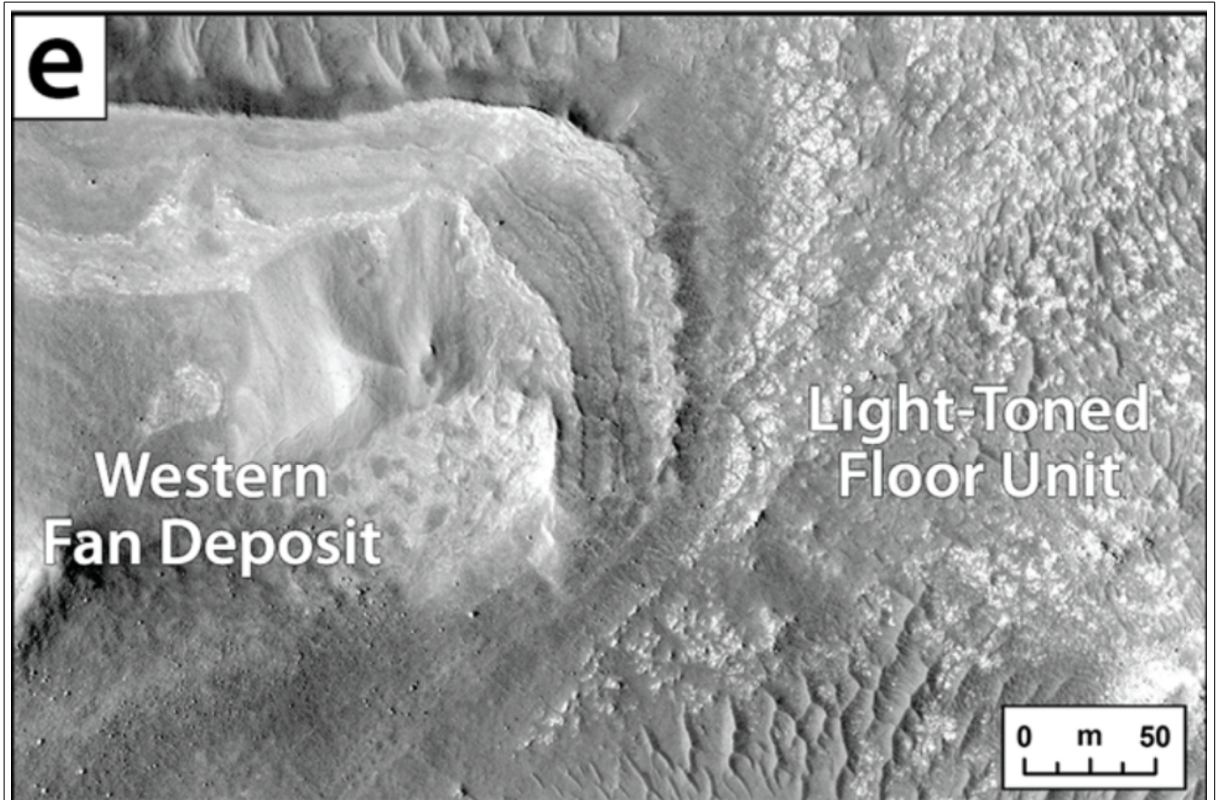

Figure 4 – Figure 9e from Goudge et al. (2015) showing the light toned floor unit (the olivine-carbonate lithology) exposed beneath the delta unit, showing 1-10 meter scale characteristic fracturing, similar to the outcrop in the Apex basalt cumulate.

***Apex Basalt as Mars analog for Olivine-carbonate lithology (light toned floor unit)***

The largest surface exposure of carbonate-bearing material identified to date was discovered in the Nili Fossae region using the CRISM instrument on the MRO spacecraft (Ehlmann et al., 2008). Brown et al. (2010) argued that the serpentinised olivine ultramafic cumulate layer at the North Pole Dome is a good analog for the olivine-carbonate lithology present in Nili Fossae and at Jezero Crater, the landing site of the Mars2020 rover. They suggested that although the olivine has been intensely replaced by the serpentine and talc, in a Martian environment, the process of partial serpentinization still gives a good match for the spectral evidence that has been detected by the CRISM instrument.

The Martian olivine-carbonate lithology shows variable carbonation and evidence for a hydration event in the form of hydroxyl bands which have been interpreted as due to smectite (Bishop et al., 2008), chlorite (Viviano et al., 2013), saponite (Ehlmann et al., 2009), serpentine (Ehlmann et al., 2009; Brown et al., 2010; Amador et al., 2018) and/or talc (Brown et al., 2010; Viviano et al., 2013), will help reveal the alteration conditions and temperature and pressure conditions that accompanied the hydration event, and may determine whether it was associated with a serpentinization, sedimentary or leaching process. The formation conditions are crucial because the presence of talc-carbonate resulting from the carbonatization of serpentine has been examined in Earth analogs in terrestrial greenstone belts such as the Pilbara in Western Australia (Brown et al., 2005, 2006), where talc-bearing komatiite cumulate units of the Dresser Formation overlie the siliceous, stromatolite-bearing Strelley Pool Chert unit (Allwood et al., 2006; Van Kranendonk et al., 2008). An in situ investigation of the Mg/Fe-phyllosilicate mineralogy and the nature of the hydration event is therefore a critical task in understanding the astrobiological potential of the carbonate and phyllosilicate deposits at Nili Fossae.